\documentclass[twocolumn,prl,aps]{revtex4}
\usepackage{amsmath}
\usepackage{amssymb}
\usepackage{graphicx}
\usepackage{dcolumn}
\usepackage{graphicx}
\usepackage{dcolumn}
\usepackage{bm}

\begin{document}

\preprint{Phys.Rev.Lett.}

\title{Quantum Hall effect in n-p-n and n-2D Topological Insulator-n junctions.}
\author{G.M.Gusev,$^1$ A.D.Levin,$^1$ Z.D.Kvon,$^2$  N.N.Mikhailov,$^2$
 and S.A.Dvoretsky,$^{2}$}

\affiliation{$^1$Instituto de F\'{\i}sica da Universidade de S\~ao
Paulo, 135960-170, S\~ao Paulo, SP, Brazil}
\affiliation{$^2$Institute of Semiconductor Physics, Novosibirsk
630090, Russia}

\date{\today}
\begin{abstract}
We have studied quantized transport in HgTe wells with inverted band
structure corresponding to the two-dimensional topological insulator
phase (2D TI) with locally-controlled density allowing n-p-n and
n-2D TI-n junctions. The resistance reveals the fractional plateau
$2h/e^{2}$ in n-p-n regime in the presence of the strong
perpendicular magnetic field. We found that in n-2D TI-n regime the
plateaux in resistance in not universal and results from the edge
state equilibration at the interface between chiral and helical edge
modes.  We provided the simple model describing the resistance
quantization in n-2D TI-n regime.


\end{abstract}

\maketitle

Recently interest in the edge state transport in the integer
quantum Hall effect (QHE) has been renewed due to observation of
the conductance quantization in the locally gated graphene layers in the
bipolar regime \cite{williams, huard, kim}. It has been demonstrated
that the density variation across the charge neutrality point
results in a p-n junction with interesting transport properties that
are absent in the QHE regime in the unipolar case. In particular, the
two-terminal resistance reveals fractional quantization in the
graphene p-n \cite{williams} or n-p-n \cite{huard, kim} junctions,
which has been attributed to chiral edge states equilibration at the
p-n interfaces \cite{abanin}. In general, the character of the QHE
transport in unipolar and bipolar regimes is quite different. In the
unipolar regime the edge states propagate in the same direction
(figure 1a), while in the bipolar regime the edge states
counter-circulate in the p and n regions, propagating parallel to
each other along the interface (figure 1c). The intermode scattering
across the interface in the presence of the disorder leads to interference between
channels, and conductance should exhibit
fractional quantization superimposed by universal conductance
fluctuations (UCF)\cite{abanin}.

\begin{figure}[ht!]
\includegraphics[width=8cm,clip=]{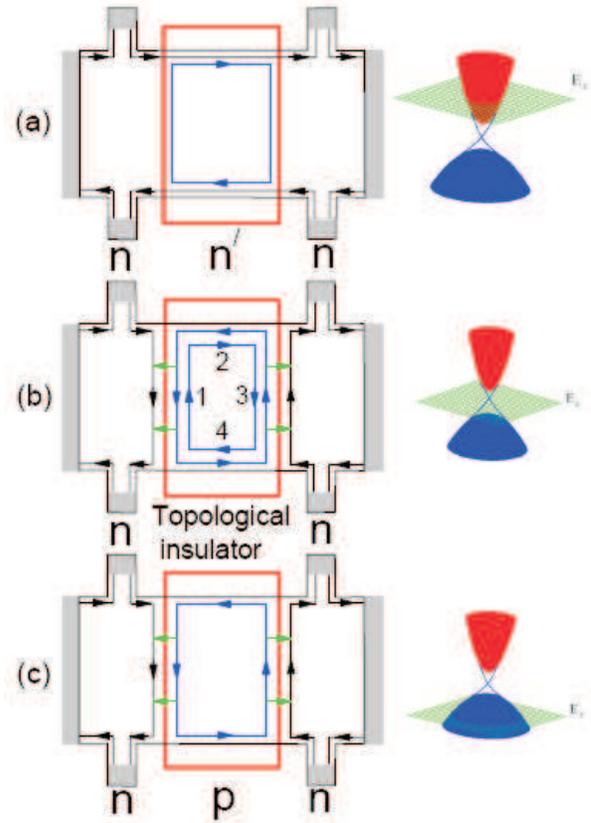}
\caption{\label{fig.1}(Color online) Schematics of edge state
propagation for different charge densities in the central local gate
region (red rectangular) and in the regions outside the local gate
in the strong magnetic field: (a) n-$n^{'}$-n junction, $n^{'}>n$,
(b) n-2DTI-n junction, (c) n-p-n junction  n=p, where n (p) refers
to negative (positive) charge density. Inserts- the energy spectrum
for different Fermi energy position in the region under local gate
at B=0.}
\end{figure}

Note, however, that UCF in the bipolar regime have not been observed,
which has been attributed to several extrinsic and intrinsic
mechanisms \cite{abanin}. Further study demonstrated that UCF in a
sufficiently small (mesoscopic) system  are robust to sample
disorder \cite{long}, but could be suppressed in the absence of the
intervalley scattering \cite{snyman}. In the latter case, the plateaux
value is expected to be shifted up or down from the quantized value,
which disagrees with experiments. Therefore, the microscopic
mechanism providing plateaux quantization in graphene p-n and p-n-p
structures still remains unclear. The semiclassical approach
\cite{carmier} confirms this conclusion.

Another interesting system, which provides for realization of the
p-n junction and study of the QHE in the bipolar regime is the
HgTe-based quantum well. The transport properties of such a system
depends on the well width $d$. In particular, when $d$ exceeds the
"critical" width approximately equal to 6.3 nm, the energy spectrum
becomes inverted and one changes to a two-dimensional topological
insulator (2D TI), or quantum spin Hall insulator phase (QSHI)
characterized by an insulating gapped phase in the bulk and
conducting edge modes, which propagate along the sample periphery
\cite{kane, bernevig, hasan, qi, konig, buhmann}. As in graphene,
both the carrier type and density in the HgTe-based well can be
electrostatically controlled via the employment of the global or
local gate \cite{gusev}. Since the ungated HgTe well is initially
n-doped, the realization of n-p or n-p-n junction requires only a
single local gate in contrast to graphene, where a combination of
the local and global control of the carrier type and density is
necessary \cite{williams, huard, kim}.

In the present paper we report the realization of local top gating
in a HgTe-based quantum well with inverted band structure for which
the density in each region could be varied across the gap, allowing
a n-p-n junction to be formed at the interfaces. Moreover, when the
Fermi energy in the region under the local gate lies in the bulk gap
band, the transport at the junction interface is described by mode
mixing between conventional QHE edge channels and pairs of
counter-propagating modes with opposite spin polarizations (figure 1
b), corresponding to QSHI. We find the fractional quantum Hall
effect plateaux $R=2\frac{h}{e^{2}}$ in the n-p-n regime in
accordance with a mode describing the counter-circulate mixing edge
state model \cite{kim, abanin}. Surprisingly, we did not find
mesoscopic conductance fluctuations, although our samples were
sufficiently small and transport would be expected to be coherent.
In the n-QHSI-n regime resistance reveals quantization close to the
$\frac{h}{e^{2}}$ value, which clearly demonstrates the existence of
the counter circulating edge states in the bulk gap region.

The $Cd_{0.65}Hg_{0.35}Te/HgTe/Cd_{0.65}Hg_{0.35}Te$ quantum wells
with (013) surface orientations and a width $d$ of 8 nm were
prepared by molecular beam epitaxy. A detailed description of the
sample structure has been given in \cite{kvon, olshanetsky}. The
six-probe Hall bar was fabricated with a lithographic length $6 \mu
m$ and width $5 \mu m$. The ohmic contacts to the two-dimensional
gas were formed by in-burning of indium. To prepare the gate, a
dielectric layer containing 100 nm $SiO_{2}$ and 200 nm
$Si_{3}Ni_{4}$ was first grown on the structure using the
plasmochemical method. Then, the TiAu gate with a width of $W= 3 \mu
m$ was deposited. Width $W$ is smaller than the distance between
potentiometric probes, therefore the voltage applied to this local
gate tunes the density only in the strip below the gate and creates
a tunable potential barrier (Insert to figure 2a). The ungated HgTe
well was initially n-doped with density $n_{s}=1.8\times10^{11}
cm^{-2}$. Several devices with the same configuration have been
studied. The density variation with gate voltage was $1.09\times
10^{15} m^{-2}V^{-1}$. For comparison, we also used a device with
the gate covering all the sample area including the potentiometric
probes, dedicated for conventional 4-probe measurements (insert to
figure 2b).
\begin{figure}[ht!]
\includegraphics[width=8cm,clip=]{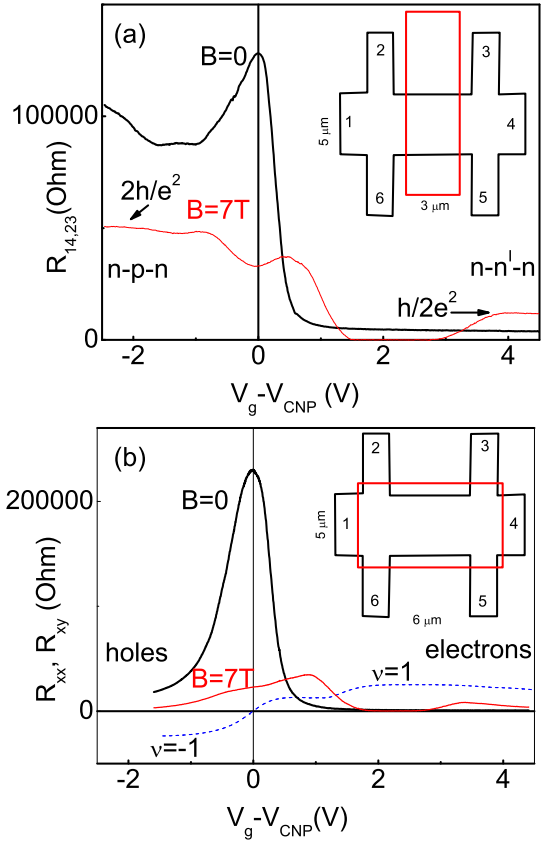}
\caption{\label{fig.2}(Color online) Color online) (a) The
longitudinal $R_{14,23}$ ( I=1,4; V=3,2) resistance as a function of
the gate voltage at zero magnetic field  (black thick line) and B=7
T (thin red line) , T=1.4 K. (b) The longitudinal $R_{xx}$ ( I=1,4;
V=3,2) resistance as a function of the gate voltage at zero magnetic
field (black thick line) and B=7 T (thin red line), and Hall
$R_{xy}$ ( I=1,4; V=2,6) resistances T=1.4 K. The insert shows
schematics view of the sample. The perimeter of the gate is shown by
rectangle.}
\end{figure}
The magnetotransport measurements in the  structures described were
performed in the temperature range 1.4-25 K and in magnetic fields
up to 12 T using a standard four point circuit with a 3-13 Hz ac
current of 0.1-10 nA through the sample, which is sufficiently low
to avoid overheating effects.

The density of the carriers in the HgTe quantum wells can be
electrically manipulated with local gate voltage $V_{g}$. The
typical dependence of the four-terminal $R_{14,23}=R_{I=1,4;V=2,3}$
resistance of one of the representative samples as a function of
$V_{g}$ is shown in Figure 2. The resistance $R_{14,23}$ in a zero
magnetic field exhibits a sharp increase when the electrochemical
potential enters the insulating bulk gap and reaches saturation at a
level that is $\sim 10$ times greater than the universal value
$h/2e^{2}$, which is expected for 2D TI phase. This value varies
from 150 to 300 kOhm in different samples. The device with a global
gate reveals a sharp peak, shown in figure 2b, when the gate voltage
induces an additional charge density, altering the quantum wells
from an n-type conductor to a p-type conductor via a QHSI state. It
has been shown \cite{konig, buhmann} that the 4-probe resistance in
an HgTe/CdTe micrometer-sized ballistic Hall bar demonstrated a
quantized plateaux $R_{14,23}\simeq h/2e^{2}$. It is expected that
the stability of the helical edge states in the topological
insulator is unaffected by the presence of a weak disorder
\cite{kane, bernevig, hasan}. Note, however, that quantized
ballistic transport has been observed only in micrometer-sized
samples and the plateaux $R_{14,23}\simeq h/2e^{2}$ is destroyed if
the sample is above a certain critical size of about a few microns
\cite{konig}. The understanding of the stability of the plateaux in
macroscopic samples requires further investigation.

The Hall effect reverses its sign and $R_{xy}\approx0$ when $R_{xx}$
approaches its maximum value (figure 2b), which can be identified as
the charge neutrality point (CNP). These behaviours resemble the
ambipolar field effect observed in graphene \cite{sarma}.
Application of the perpendicular magnetic field leads to suppression
of the peak in both structures, although the behaviour of the
resistance in the electron and hole parts of the spectrum is quite
different. One can see that the resistance in a local gate device
shows the plateaus $R_{14,23}\approx 2\frac{h}{e^{2}}$ in the n-p-n
region and $R_{14,23}\approx\frac{1}{2}\frac{h}{e^{2}}$ in the
n-$n'$-n region, while the device with a global gate demonstrates
conventional quantum Hall behaviour. Note also that $R_{14,23}=0$
near $V_{g}-V_{CNP}\approx 2 V$ in both structures. Figure 3 shows
the resistance in the local gate device in the voltage-magnetic
field plane. One can see the evolution of the longitudinal
resistance with magnetic field and density in the n-p-n, n-$n'$-n
and n-TI-n regions. The resistance peak drops dramatically in a
magnetic field above 3T and shows plateaux-like behaviour
$R_{14,23}\approx 2\frac{h}{e^{2}}$ in the $B-V_{g}$ plane in the
n-p-n region. Such resistance decrease demonstrates the transition
to the edge state transport regime.  For positive gate voltage
($V_{g}-V_{CNP} > 3.5 V$), when n-$n'$-n junctions are expected to
be formed, one can see a series of the fully-developed plateaux with
magnetic field. As B increases the final plateaux
$R_{14,23}\approx\frac{1}{2}\frac{h}{e^{2}}$ emerges. Similar
behaviour is observed around CNP, when the QHSI phase is formed
under local gate, the plateaux $R_{14,23}\approx1.3\frac{h}{e^{2}}$
develops in a wide range of the magnetic field and narrow range of
density. Slightly above CNP, in the electronic  part of the peak,
the resistance value is shifted up and approaches the value
$R_{14,23}\approx1.43\frac{h}{e^{2}}$. In the region between this
plateaux and $R_{14,23}\approx\frac{1}{2}\frac{h}{e^{2}}$, the
resistance vanishes and shows pronounced minima.
\begin{figure}[ht!]
\includegraphics[width=8cm,clip=]{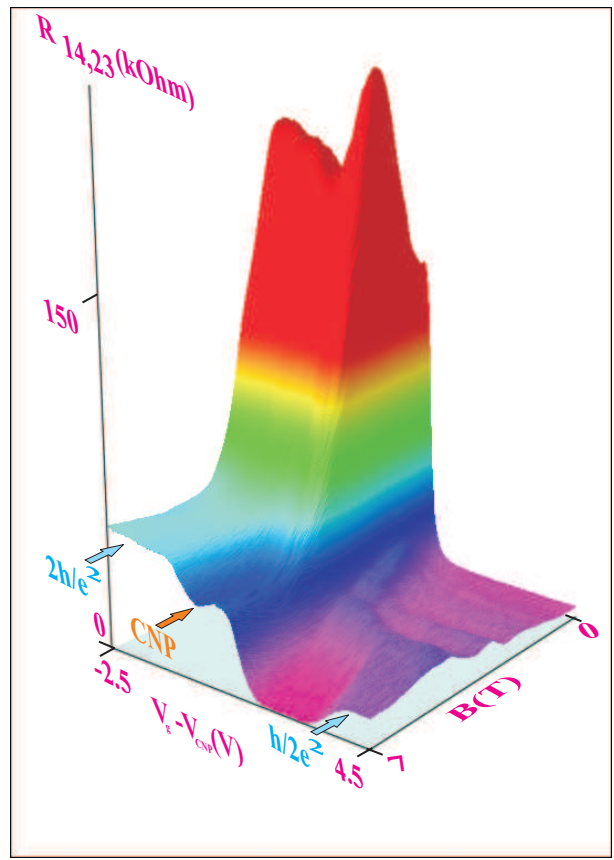}
\caption{\label{fig.3}(Color online) The longitudinal resistance
$R_{14,23}$ as a function of the gate voltage and magnetic field,
T=1.4 K. Two plateaux $R_{14,23}\approx 2\frac{h}{e^{2}}$ and
$R_{14,23}\approx\frac{1}{2}\frac{h}{e^{2}}$ are indicated by blue
arrows. CNP is indicated by orange arrow.}
\end{figure}
In the rest of the paper, we will focus on the explanation of the
resistance quantization in  HgTe quantum wells in the bipolar regime
in a strong magnetic field.

QHE edge state transport in the Hall bar geometry with a gate finger
across the device has been extensively explored in the past in the
monopolar regime \cite{haug}. The 4-probe resistance is expected to
be quantized at values
$R_{14,23}=\frac{h}{e^{2}}(\frac{1}{\nu}-\frac{1}{\nu_{g}})$, where
$\nu_{g}$ is the Landau level  (LL) filling factor in the gate
region, and $\nu$ is the filling factor in the region outside of the
gate. Indeed this formula perfectly describes the resistance
behaviour at $V_{g}-V_{CNP}>0$ in a magnetic field above 3T,
notably. $R_{14,23}=\frac{h}{e^{2}}(\frac{1}{1}-\frac{1}{1})=0$ and
$R_{14,23}=\frac{h}{e^{2}}(\frac{1}{1}-\frac{1}{2})=\frac{1}{2}\frac{h}{e^{2}}$.

In the bipolar regime, an unusual fractional resistance plateaux
arises from the equilibration between counter circulating edge
states in the p and n regions  ( see figure 1c) \cite{abanin}. In
the 2-probe configuration, the net resistance is described by three
quantum resistors in series:
$R_{2T}=\frac{h}{e^{2}}(\frac{1}{\nu}+\frac{1}{\nu_{g}}+\frac{1}{\nu})=\frac{h}{e^{2}}\frac{2\nu
+\nu_{g}}{\nu_{g}\nu}$, which indeed has been observed in graphene
n-p-n junctions \cite{kim}. The quantization of the 4-probe
resistance is given by slightly different equation
$R_{4T}=R_{14,23}=\frac{h}{e^{2}}(\frac{1}{\nu}+\frac{1}{\nu_{g}})$.
This formula agrees with our observation of plateaux
$R_{14,23}=\frac{h}{e^{2}}(\frac{1}{1}+\frac{1}{1})=2\frac{h}{e^{2}}$
in the n-p-n regime (figure 2a and figure 3). It is worth noting
that in graphene it is difficult to obtain the plateau at this
value, since the valley and spin splitting is small and
${\nu}=\nu_{g}=\pm 2,\pm 6.....$. The advantage of the graphene
structure is the sharper (but not abrupt) potential step on the
scale of the magnetic length \cite{kim}.

A more interesting situation occurs when the Fermi energy in the
region under the local gate tunes through the 2D TI phase (figure
1b). Such a situation allows us to use QHE mode propagation for
investigation of the intrinsic transport characteristics of the
topological insulator. As mentioned above, the gaplessness of the
edge states in TI is protected against time reversal symmetry (TRS),
which must result in the robust ballistic transport. However, a
magnetic field perpendicular to the 2D layer breaks the TRS and
thereby enables elastic scattering between counter propagating
chiral edge states. A number of different conflicting scenarios has
been developed for TRS breaking in the QHSI system \cite{konig,
tkachov, maciejko}. The more realistic  models \cite{tkachov,
maciejko} have demonstrated that the counter-propagating helical
edge states persist in a strong B. The magnetic field does not
affect the gap but it modifies the energy spectrum of the edge
states and generates backscattering between the counter propagating
modes \cite{tkachov}. One of the helical modes propagate in the same
direction as the edge state outside of the gate, while the other has
the opposite direction and, therefore, flows parallel to the outside
mode (figure 1 b).

The edge modes are described by the local chemical potentials $\xi,
\varphi$ and $\psi$, where $\varphi$ and $\psi$ are electrochemical
potentials for two spin states $\uparrow$ and $\downarrow$ in the
region under local gate propagating along interface 3 and 4 (figure
1b) and along edges of the sample 1 and 2 (figure 1b), and $\xi$ is
potential outside of the gate region propagating along lines 3 and 4
in figure 1b in the same direction, as mode $\varphi$. We can
introduce phenomenological constant $\gamma$, which represent spin
flip scattering between modes $\varphi$ and $\psi$. The scattering
between conventional QHE edge mode $\xi$ and helical modes $\varphi$
and $\psi$ is characterized by 2 parameters $\lambda$ and $\beta$
consequently. The edge state transport can be described by equations
for particle density \cite{abanin2, dolgopolov}, taking into account
the scattering between edge modes
\begin{eqnarray}
\partial_{x}\varphi_{i}=\gamma(\varphi_{i}-\psi_{i}),\\
-\partial_{x}\psi_{i}=\gamma(\psi_{i}-\varphi_{i}), i=2,4
\end{eqnarray}
The equations describe the variables $\xi_{i}$ at the edges 1 and 3:
\begin{eqnarray}
\partial_{y}\xi_{i}=\lambda(\xi_{i}-\varphi_{i})+\beta(\xi_{i}-\psi_{i}),i=1,3
\end{eqnarray}
Similar couple of equations describe the potentials $\varphi_{1,3}$
and $\psi_{1,3}$ at the edges 1 and 3. The system of 10 differential
equations have been solved numerically, and we found the current as
$I=\frac{e^{2}}{h}\sum \mu_{i}$, where $\mu_{i}$ is the local
electrochemical potential near the opposite edges of the Hall bar.
The described transport model reproduces the near quantized value of
the resistance $R_{14,23}\approx1.3\frac{h}{e^{2}}$ with 3
parameters $\lambda=0.8\mu m^{-1}$ and $\gamma=0.27 \mu m$ and
$\beta=0.5 \mu m^{-1}$. Indeed our model correctly reproduces the
value of the resistance $R_{14,23}\approx 2\frac{h}{e^{2}}$ for
$\gamma\approx 0$, which corresponds to the n-p-n situation (figure
1c). We note that the parameter $\gamma$ can be independently
obtained from the measurements in global gate sample. The 4-terminal
resistance is not universal in the presence of the backscattering
and can be described by transport equations \cite{abanin2} similar
to 1-3. We performed numerical calculations for  counter-propagating
potentials in global gate sample and found resistance. Our
calculations reproduced the value $R_{xx}\approx h/e^{2}$, which has
been observed in experiment at B=7 T (figure 2b) with parameter
$\gamma\approx 0.27 \mu m^{-1}$ in agreement with previous
situation. The backscattering mean free path $l=\lambda^{-1} \approx
0.1 \mu m$ for parallel current-currying states seems reasonable,
and is taken for the mode with the same polarization. Indeed the
scattering between counter propagating modes is much weaker.
Therefore, the plateaux-like behaviour of the resistance
unambiguously demonstrates the existence of the counter circulating
modes, when the Fermi level tunes through the bulk gap. Our
observation provides considerable support for the models
\cite{tkachov, maciejko}, which predicts persistence of the helical
modes in the strong magnetic field. In our model the resistance
value exceeds $h/e^{2}$ for any the parameters of the backscattering
between modes. It is worth emphasizing that the transport model in
the monopolar regime predicts the resistance quantization in
fractions smaller than 1.

We also measured the temperature dependence of the resistance in all
regimes and found that all plateaus remain unchanged when the
temperature decreases from 10 to 1.4 K. In the coherent regime for
small enough samples the theory \cite{abanin} predicts UCF in
bipolar structures. We expected the coherence length in our samples
to be of the order of $\sim 1\mu m$ at 1.4 K \cite{olshanetsky2}
and, therefore, UCF in our samples could be not completely
suppressed. However, similarly to the graphene n-p-n junction, we
did not observe UCF as a function of Fermi energy or magnetic field
either in n-p-n or in n-2D TI-n regimes.

In conclusion, we have studied the transport properties of the
HgTe-based quantum well with inverted band structure with gate
finger across the Hall bar geometry in a strong magnetic field. The
narrow gap band structure of HgTe allows local gate field control of
the  carrier type and density, and hence, the creation of a bipolar
n-p-n junction within a single quantum well. We observed a
fractional resistance plateau $2h/e^{2}$ originating from the mixing
of modes at the p-n interfaces. By varying the voltage on the local
gate, we studied the QHE transport in the n-2D TI-n regime, where
two counter propagating helical modes circulate along the junction
interface. The intermode scattering between helical states and the
QHE chiral edge mode results in resistance quantization at a value
$R_{14,23}\approx1.3\frac{h}{e^{2}}$. This value cannot be explained
by the transport model in the monopolar regime. This effect can be
used to explore the backscattering mechanism in a 2D topological
insulator. Our observations support the model that predicts the
robustness of helical modes in the presence of a perpendicular
magnetic field \cite{tkachov}.

We thank O.E.Raichev for helpful discussions. The financial support of
this work by FAPESP, CNPq (Brazilian agencies), RFBI and RAS
programs "Fundamental researches in nanotechnology and
nanomaterials" and "Condensed matter quantum physics" is
acknowledged.

\end{document}